\documentclass[a4paper,10pt]{article}
\usepackage{amsmath,amssymb,epsfig,graphicx} 
\usepackage{setspace}

\begin{document}

\onehalfspacing
%\doublespacing

\title{GR@PPA 2.8.3 update}

\author{Shigeru Odaka\footnote{E-mail: \texttt{shigeru.odaka@kek.jp}}\\
High Energy Accelerator Research Organization (KEK)\\
1-1 Oho, Tsukuba, Ibaraki 305-0801, Japan}

\date{}

\maketitle

\begin{abstract}
GR@PPA 2.8 is a program package including event generators for single and 
double weak-boson production processes at hadron collisions, 
in which a jet matching method is implemented for simulating the weak-boson 
kinematics in the entire phase space.
Since the initial release in November, 2010,
several improvements have been applied to the program components.
This report describes the improvements and changes applied so far, 
up to the 2.8.3 release.
\end{abstract}

\section{Introduction}
\label{sec:intro}

GR@PPA 2.8~\cite{Odaka:2011hc}, 
a program package including event generators for single and 
double weak-boson production processes at hadron collisions, 
has been released in November, 2010.
The included event generators support an initial-state jet matching method 
which allows us to simulate the weak-boson productions 
in the entire phase space.
The package includes custom-made parton shower (PS) routines 
in order to ensure a satisfactory performance of the matching method.
It has been demonstrated that the simulation result reproduces the transverse 
momentum ($p_{T}$) spectrum of $Z$ bosons produced at FNAL Tevatron 
with a very good precision.

Since the initial release of GR@PPA 2.8, together with some bug fixes, 
we have applied several improvements to the program components.
The improved versions have been released as updates 
on the GR@PPA Web page~\cite{Web:grappa}.
In this report, we describe the improvements and changes that we have applied 
until the 2.8.3 update.

%--------1---------2---------3---------4---------5---------6---------7---------8
\section{2.8.1 and 2.8.2 updates}
\label{sec:pre283}

The modifications in the 2.8.1 and 2.8.2 updates were rather minor 
because they were released mostly for bug fixes.
In the 2.8.1 update, 
we increased the lower cutoff ($Q_{0}$) of built-in parton showers (PS)
from 4.6 GeV to 5.0 GeV.
Our PS always assumes the contribution of massless five flavors of quarks. 
Thus, the $Q_{0}$ value must be larger than the $b$-quark mass.
The original value of 4.6 GeV is safe for the standard value of the $b$-quark 
mass.
However, in some PDF sets, 
the $b$-quark mass is assumed to be larger than the standard value 
for investigating the mass effect.
We have increased the $Q_{0}$ value in order to avoid unexpected behaviors 
of PS routines when the internal $Q$ value falls below the threshold.

Another modification is relevant to the definition of the energy scale.
In our matching method, 
the simulation of divergent leading-logarithmic (LL) terms are shared between 
the parton shower (PS) and the matrix elements (MEs) of radiative (1-jet) 
processes.
The initial-state PS applied to non-radiative (0-jet) events simulates 
small-$Q^{2}$ radiations, 
and 1-jet MEs explicitly simulate large-$Q^{2}$ hard radiations.
We use the factorization scale ($\mu_{F}$) as the boundary of this sharing.
Since this boundary is usually in a visible region, it is crucial to define 
the $\mu_{F}$ value consistently in 0-jet and 1-jet event generations
in order to accomplish a precise matching.

In the 2.8.2 update, 
a modification has been applied so that the $\mu_{F}$ of the corresponding 
0-jet subprocess can be referred to from routines for calculating 1-jet MEs.
This improvement guarantees a consistent definition of the boundary.
For this purpose, we have added a subroutine {\tt setscale} to the file 
{\tt grcpar.F} in sample programs.
This subroutine is called in the subroutine {\tt GRCUSRSETQ} to determine 
the $\mu_{F}$ value of the event to be generated, 
and in 1-jet ME routines to determine the upper bound of the LL subtraction.
Note that {\tt GRCUSRSETQ} is called only when the "user-defined energy scale" 
option is chosen.
Thus, for the 2.8.2 update and later, 
we recommend users to always use this option by setting {\tt ICOUP = 6} 
and {\tt IFACT = 6} in the subroutine {\tt GRCPAR}, 
and appropriately define the energy scales in {\tt setscale} 
if they want to utilize the matching method.

Random number generators (RNGs) were replaced in the 2.8.2 update.
Two kinds of RNGs are used in the GR@PPA event generators: 
one for the generation of hard interaction processes in the framework of 
BASES/SPRING and another for PS routines.
It has been found that originally used RNGs do not have  
sufficient performances for a large scale event generation; 
a looping was found.
In the 2.8.2 update, the RNG for the hard-interaction generation was 
replaced with a Fortran version of the Mersenne Twister method~\cite{mt19937}, 
and that for the PS routines with a program of Chandler~\cite{randgen}.
The usage of the RNG has also been improved in the hard-interaction generation.

%--------1---------2---------3---------4---------5---------6---------7---------8
\section{2.8.3 update}
\label{sec:283update}

Two important improvements were applied in the 2.8.3 update.
The first one is concerning the kinematics of the final-state PS.
Since PS branches always have certain transverse momenta, 
the application of PS necessarily adds non-zero invariant masses 
("jet mass") to partons.
Thus, it increases the energy of the final-state system.
Originally, we adjusted (decreased) the momenta of the final-state particles 
after completing the PS 
in order to compensate for the increase of the total energy.
The adjustment had to be applied not only to the showered partons but also to 
weak bosons in order to preserve the momentum balance.
However, the implementation of the final-state PS depends on the maximum 
energy scale ($\mu_{\rm FSR}$) which can be given arbitrarily.
Namely, rigorously to say, the momentum of weak bosons is dependent 
on the choice of $\mu_{\rm FSR}$ in this model.
Although usually the dependence is not very significant, 
it is unreasonable that a visible quantity depends on the choice of 
an arbitrary parameter.

This momentum adjustment was removed in the 2.8.3 update.
Instead, the momenta of initial-state partons are adjusted (increased) 
to make the initial- and final-state energies become consistent.
People may worry that such an adjustment may violate the consistency 
with the parton distribution function (PDF).
However, the initial-state parton momentum is not a physical quantity.
It is "derived" from the quantities of visible particles, such as electrons 
in $ep$ collision experiments, by ignoring the jet mass.
We suppose that the adjustment in the initial state must be more reasonable 
than that in the final state.

\begin{figure}
\begin{center}
\includegraphics[scale=0.65]{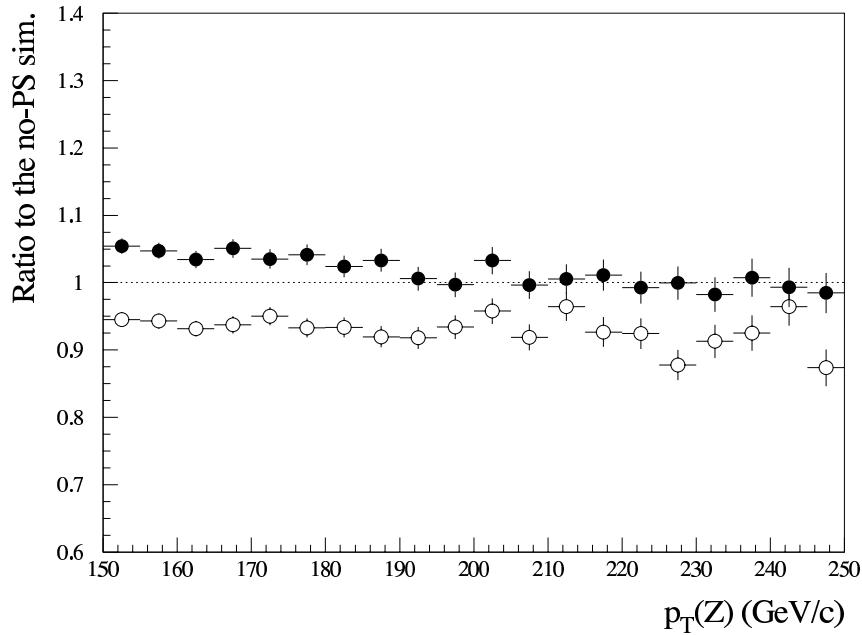}
\caption{\label{fig:FSRkinem_ratio}
effect of the modification in the final-state PS kinematics.
The $p_{T}$ spectrum of $Z$-boson production at high-$p_{T}$ has been 
simulated in the 14-TeV LHC condition, 
based on the matrix elements for $Z$ + 1 jet production processes.
The ratio of the simulated spectrum to the direct prediction 
from the matrix elements is plotted.
The results from the 2.8.2 (open circles) and 2.8.3 (filled circles) 
updates are compared.
}
\end{center}
\end{figure}

Figure~\ref{fig:FSRkinem_ratio} shows the effect of this modification.
We have generated $Z$ + 1 jet events in the LHC condition 
with a center-of-mass (cm) energy of 14 TeV.
The transverse momentum ($p_{T}$) spectrum of $Z$ bosons at high $p_{T}$ 
should be reproduced by this simulation.
The PS simulations of GR@PPA have been fully applied 
with all the energy scales set to the $Z$-boson mass 
and, as usual, lower-$Q^{2}$ effects have been simulated 
by using PYTHIA~\cite{Sjostrand:2006za} down to the hadron level.
The ratio of the simulated result to the direct prediction from 
the matrix elements is plotted in the figure 
in order to have a small difference visible.
The results from the 2.8.2 (open circles) and 2.8.3 (filled circles) 
updates are plotted.
We can see that the modification applied in 2.8.3 enhances the spectrum 
in this high-$p_{T}$ region nearly constantly by about 10\%.

Concerning the final-state PS, 
some minor modifications have also been applied.
In the original 2.8 release, while the default was $\mu_{\rm FSR} = \mu_{F}$,
the $\mu_{\rm FSR}$ value was set to the $p_{T}$ of the showering parton 
when the $p_{T}$ was smaller than $\mu_{F}$.
This exception was introduced because it seemed to be unnatural that 
partons could radiate very hard secondary partons having transverse 
momenta larger than the parent-parton momenta.
However, such very hard branches are naturally suppressed by the 
"forward" condition in our PS kinematics.
Because we suppose that exceptions should be minimized, 
this exception was removed in the 2.8.3 update. 
The $\mu_{\rm FSR}$ value is always set to $\mu_{F}$ since the 2.8.3 update.
A similar modification has also been made in the implementation of 
the final-state PS for partons radiated in the initial-state PS.
Originally, the energy scale of this PS was set to the $p_{T}$ of the branch 
in which the parton of interest was produced.
The definition was changed to the $Q$ value of the production branch 
in the 2.8.3 update. 
In any case, these are very details of the model 
and there is no strong reason supporting these changes.
Hence, these details may be modified again in future updates.

Another marked improvement in 2.8.3 is concerning the choice of 
$\Lambda_{\rm QCD}$ for parton showers.
All the PS routines in GR@PPA are using the 1-loop approximation 
of $\alpha_{s}$ with the leading-order (LO) splitting functions.
Thus, respecting a strict theoretical consistency, 
we can use only those PDFs which are based on the same approximation.
However, from technical aspects, 
we can use other PDFs based on different approximations 
by applying a backward-evolution PS in the initial state.
We provide such a PS (QCDPSb) in the GR@PPA 2.8 package.
The choice of $\Lambda_{\rm QCD}$ is not trivial in such usage.

In the original GR@PPA 2.8, 
the default $\Lambda_{\rm QCD}$ value (0.165 GeV) was used when any mismatch 
to the selected PDF was found in the $\alpha_{s}(Q^{2})$ definition.
This default is the value from CTEQ6L1~\cite{Pumplin:2002vw}, 
a LO PDF with the 1-loop $\alpha_{s}$, 
which corresponds to $\alpha_{s}(m_{Z}^{2}) = 0.130$.
Note that the world average of $\alpha_{s}(m_{Z}^{2})$ is 0.118. 
The $\alpha_{s}$ value is increased in CTEQ6L1 
in order to reproduce the measured rapid evolution at low $Q^{2}$ 
($Q$ = a few GeV).
However, the initial-state PS in GR@PPA covers the energy region of 
$Q > 5 {\rm ~GeV}$ only. 
This coverage is sufficient for simulating the perturbative part 
of the initial-state QCD activity for hard processes 
such as weak-boson production.
We should use $\alpha_{s}$ as close to the values in the selected PDF 
as possible in this region, in order to achieve a consistent evolution.
The application of the CTEQ6L1 value must not be appropriate 
when the 2-loop $\alpha_{s}$ is assumed in the selected PDF.

Note that the $Q^{2}$ dependence is not very different between the 1-loop 
and 2-loop $\alpha_{s}$ formulae at $Q > 5 {\rm ~GeV}$.
The difference is only 5\% even at $Q = 5 {\rm ~GeV}$ 
when we choose an identical fixing condition of 
$\alpha_{s}(m_{Z}^{2}) = 0.118$.
On the other hand, 
the 1-loop $\alpha_{s}$ with $\Lambda_{\rm QCD} = 0.165 {\rm ~GeV}$ is 
almost always 10\% larger than the 2-loop value giving 
$\alpha_{s}(m_{Z}^{2}) = 0.118$ in the range of interest.
Since the GR@PPA 2.8.3 update, 
in order to achieve a more precise matching in the practical range, 
we set the $\Lambda_{\rm QCD}$ for the parton showers 
so that the derived 1-loop $\alpha_{s}(Q^{2})$ matches the value 
of the selected PDF at $Q = 10 {\rm ~GeV}$, 
regardless to the definition of $\alpha_{s}(Q^{2})$ in the PDF.
This definition gives $\alpha_{s}(Q^{2})$ within $\pm 2\%$ from the values 
in PDF in a practical range, $5 < Q < 100 {\rm ~GeV}$, 
and only a 2.3\% deviation even at $Q = 250 {\rm ~GeV}$, 
even when the PDF uses the 2-loop $\alpha_{s}$.

\begin{figure}
\begin{center}
\includegraphics[scale=0.65]{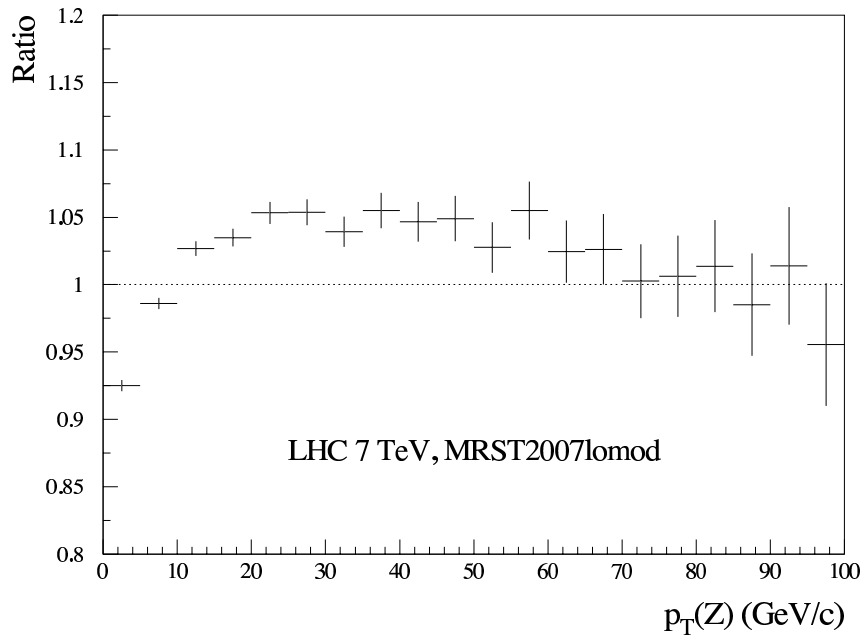}
\caption{\label{fig:lambdaTest}
effect of a different choice of the $\Lambda_{\rm QCD}$ parameter 
in the initial-state parton shower.
The normalized $p_{T}$ distribution of $Z$ bosons, 
$(1/\sigma)d\sigma/dp_{T}$, has been simulated in the 7-TeV LHC condition, 
using MRST2007lomod for PDF with the backward evolution.
The ratio of the result with the original GR@PPA 2.8 default, 
$\Lambda_{\rm QCD} = 0.165 {\rm ~GeV}$, 
to that with the GR@PPA 2.8.3 default is plotted.
}
\end{center}
\end{figure}

Figure~\ref{fig:lambdaTest} shows the effect of this change 
in the $\Lambda_{\rm QCD}$ definition.
The normalized $p_{T}$ distribution of $Z$ bosons, $(1/\sigma)d\sigma/dp_{T}$, 
has been simulated in the LHC condition with a cm energy of 7 TeV, 
by using the MRST2007lomod~\cite{Sherstnev:2007nd} PDF 
with the backward-evolution PS (QCDPSb).
This PDF is a so-called modified leading-order (LO*) PDF 
using the 2-loop $\alpha_{s}$ for the evolution.
The ratio of the result with $\Lambda_{\rm QCD}$ = 0.165 GeV to that 
with the 2.8.3 default is plotted.
We can see a nearly constant difference between the two simulations 
at the level of 5\% at medium $p_{T}$. 
The spectrum at medium $p_{T}$ is predominantly determined by 
the multiple QCD-radiation effect simulated by PS~\cite{Odaka:2009qf}.
This effect is enhanced in the original simulation 
due to the application of a larger $\alpha_{s}$ value.
The deficit at small $p_{T}$ ($< 10$ GeV/$c$) compensates for this enhancement 
because the overall normalization is the same. 

In addition to the above two improvements, 
a process-dependent K-factor is supported since the 2.8.3 update.
The cross section of each process is multiplied by a parameter {\tt fnorm} 
which can be specified in the subroutine {\tt GRCPAR} provided in the file 
{\tt grcpar.F} of sample programs.
This parameter has to be defined for all activated processes 
if non-unity value is specified for any process.
This option may be useful for testing the effect of relative normalization 
uncertainties between activated processes.

\section{Summary}
\label{sec:summary}

Since the initial release of the GR@PPA 2.8, together with some bug fixes, 
we have applied several improvements to the program components.
The most remarkable change up to the GR@PPA 2.8.3 update 
is concerning the final-state parton shower (PS) kinematics.
We have modified the kinematics so that the application of PS 
should never alter the momenta of other particles in the final state.
Another important improvement is concerning the definition of $\Lambda_{\rm QCD}$ 
for parton showers.
It is technically possible to use PDFs based on $\alpha_{s}$ approximations 
different from that in PS by adopting a backward-evolution PS.
Since the 2.8.3 update, the $\Lambda_{\rm QCD}$ value is chosen so that 
the derived $\alpha_{s}(Q^{2})$ should reproduce the value in the selected PDF 
as precisely as possible in such applications.
These changes result in alternations at the level of as much as 10\% 
in the $Z$-boson $p_{T}$ spectrum.
It will soon become possible to experimentally verify such details 
by utilizing high-precision data from LHC.



%% References


%\bibliographystyle{elsarticle-num}
%\bibliography{grappa28}



\end{document}